\documentclass[12pt]{iopart}

\begin{document}

\title[The quest for novel modes of excitation in exotic nuclei]{The quest for novel modes of excitation in exotic nuclei}

\author{N. Paar}
\address{Physics Department, Faculty of Science, University of Zagreb,
Bijeni\v cka 32, 10000 Zagreb, Croatia}
\ead{npaar@phy.hr}
\begin{abstract}
This article provides an insight into several open problems in the quest for novel modes
of excitation in nuclei with isospin asymmetry, deformation and finite temperature characteristic
in stellar environment. 
Major unsolved problems include the nature of pygmy dipole resonances, 
the quest for various multipole and
spin-isospin excitations both in neutron-rich
and proton drip-line nuclei mainly driven by loosely bound nucleons, excitations in 
unstable deformed nuclei and evolution of their properties with the shape
phase transition. Exotic modes of excitation in nuclei at finite temperatures characteristic
for supernova evolution present open problems with possible impact in modeling
astrophysically relevant weak interaction rates. 
All these issues challenge self-consistent many body theory frameworks at the
frontiers of on-going research, 
including nuclear energy density functionals, both phenomenological and constrained 
by the strong interaction physics of QCD, models based on low-momentum two-nucleon
interaction  $V_{low-k}$ and correlated realistic nucleon-nucleon interaction $V_{UCOM}$, 
supplemented by three-body force, 
as well as two-nucleon and three-nucleon interactions derived from the chiral 
effective field theory.
Joined theoretical and experimental efforts, including research
with radioactive isotope beams, are needed to provide insight into dynamical
properties of nuclei away from the valley of stability, involving the interplay of isospin asymmetry, 
deformation and finite temperature.
\end{abstract}

\maketitle

\section{Introduction}

Nuclei away from the valley of stability are characterized by unique structure
properties including weak binding of outermost nucleons, coupling between bound
states and the particle continuum, diffuse neutron density distributions, formation of
neutron skin and halo structures. Due to these properties, the structure of multipole
response in unstable nuclei can be modified and novel exotic modes of excitation
may appear. In particular, the dipole response of neutron-rich nuclei is characterized
by the fragmentation of the strength distribution and its spreading into the low-energy region,
and by the mixing of isoscalar and isovector components~\cite{PVKC.07}. While in light nuclei the onset
of dipole strength in the low-energy region is caused by non-resonant independent single-particle
excitations of the loosely bound neutrons, several theoretical analyses have predicted
the existence of the pygmy dipole resonance (PDR) in medium-mass and heavy nuclei, i.e. the
resonant oscillation of the weakly-bound neutron skin against the isospin saturated protonÐ
neutron core. Although the properties of the PDR have been studied with numerous
many body methods and effective nuclear interactions, there is a considerable
number of open questions about its dynamics that are at present time under discussion. 

The richness of the structure properties and numerous challenging results 
about the nature of PDR open perspectives to develop a whole new area of
research on other exotic modes of excitation in weakly bound nuclei. One 
could expect, at least in principle, that the structure properties of exotic nuclei
could result in novel modes of excitation of various multipolarities,  in charge-exchange
channel and nuclei with deformation, as well as in nuclei at finite temperatures 
characteristic for the stellar environment and supernova evolution.

Exotic modes of excitation probe several aspects of the 
underlying structure properties and effective nuclear interactions, including the in-medium modification of
the nucleon-nucleon interaction, and the interplay between different degrees of freedom in
dissipative processes that determine the damping mechanism. These modes could also
provide sensitive test of novel microscopic theory frameworks based on nuclear energy density
functional~\cite{Ben.03,Dob.07,Toi.08,Kor.08,CB.08,Fin.06,Nik.08,RL.09} or
low-momentum two-nucleon (NN) interactions,  $V_{low-k}$ based on renormalisation group
transformations~\cite{Bog.03} or $V_{UCOM}$ obtained by the unitary correlation operator
method~\cite{Fel.98}.
$V_{low-k}$  has also been supplemented with a three-body force (NNN)~\cite{Bog.09}. 
New developments in connecting the QCD with low-energy 
nuclear physics through chiral effective field theory ($\chi$EFT), which resulted in
consistent derivation of NN and NNN forces, could also provide challenging
theory framework to probe the structure properties of exotic modes~\cite{Epe.09}.
Progress in theoretical predictions of
exotic modes has also initiated a number of experimental studies on low-energy multipole
response, including for instance scattering of photons, protons, and alpha particles on nuclei, 
as well as excitations involving radioactive isotope beams (e.g. Refs.~\cite{Adr.05} -\cite{Kli.07}).

In the following sections
an overview is given on several challenging research topics 
on exotic modes of excitation in nuclei, that are not understood at present time, 
and could be addressed in the future by theory frameworks either based on modern energy
density functionals or low-momentum NN interactions supplemented
by NNN forces. All these research topics necessitate strong
support from the experiment,  in order to validate theoretical methods, findings, 
and implementation of various effective interactions in modeling the 
nuclear many-body problem under conditions of pronounced isospin asymmetry, 
deformation, and finite temperature.

\section{Self-consistent theory of excitations in exotic nuclei}

One of the central aims in nuclear physics is the construction of effective
interactions suitable for self-consistent solving the nuclear many-body problem, 
either within the nuclear energy density functional theory or models based on realistic NN and NNN interactions.
In the former case, the parameters of effective interactions are usually constrained
at the level of the self-consistent mean field model by selection of data on the ground state
properties of atomic nuclei and nuclear matter. The most prominent self-consistent 
mean-field models for describing nuclear structure and low-energy dynamics
are based on Skyrme energy functional, finite range Gogny interaction, and 
relativistic energy density functional~\cite{Ben.03}. More recently, these frameworks also
provided insight  into various aspects of nuclear structure in unstable nuclei, 
including exotic modes of excitation in neutron-rich nuclei, e.g. pygmy dipole 
resonance~\cite{PVKC.07}, di-neutron vibration mode~\cite{Mat.05}, 
low-energy quadrupole response~\cite{Kha.02,Miz.09}, etc. The self-consistent RPA 
with Gogny interaction, using the Gaussian expansion method which includes coupling
to the continuum enables studies of low-energy multipole excitations in drip-line 
nuclei~\cite{Nak.09}.

Excitations involving weakly bound nucleons represent
very fine structure phenomena, and could be used as sensitive benchmarks
in construction of novel effective nuclear interactions and many body methods.
For a microscopic description of transition spectra spreading widths, of particular
importance are the effects due to coupling to complex configurations. In a 
recent formulation of relativistic quasiparticle time blocking approximation,
coupling of two-quasiparticle excitations to collective vibrations is included within a fully 
consistent calculation scheme based on covariant energy density 
functional theory~\cite{Lit.08,Lit.09}. 
Another approach which takes into account complex configurations, second RPA (SRPA),
is based on extension of the RPA which, in addition to $ph$ configurations, also includes
$2p2h$ excitations and residual interaction terms that couple $ph$ and $2p2h$ 
configurations, as well as $2p2h$ configurations among themselves. However, due to
numerical difficulties, the SRPA equations have usually been solved by reducing to
Tamm-Dancoff approach, the residual interaction terms that couple $2p2h$ with $2p2h$
configurations have been neglected,  severe restrictions on the configuration space have been imposed,
and there have been no consistency between the ground state description and residual SRPA
interaction terms. Only very recently, self-consistent SRPA have been established,
based on correlated Argonne V18~\cite{Pap.09} and Skyrme effective interactions~\cite{Gam.10},
and employed in the initial studies of giant resonances and low-lying $0^+$ and $2^+$ states.
Future developments and studies based on SRPA could provide crucial information about 
excitations in unstable nuclei and their spreading widths.

At present time exotic modes of excitation are still out of reach for theory frameworks
starting from the strong interaction physics of QCD. One promising approach could be
the nuclear energy density functional theory (NEDF), based on chiral dynamics and the symmetry breaking pattern 
of low-energy QCD~\cite{Fin.06}, and recently extended to the description of collective nuclear
excitations~\cite{Fin.07}. The proton-neutron random phase approximation, based on relativistic point-coupling
Lagrangian, has been applied to investigate the role of chiral pion-nucleon dynamics in excitation
modes involving spin and isospin degrees of freedom, e.g. isobaric analog states and Gamow-Teller 
resonances. The NEDF framework could be further extended for studies of other
excitation modes of interest. 
Nuclear potentials have also been constructed in another approach based 
on effective field theories, that respect the symmetry pattern of QCD 
and produce observables in a systematic and controlled expansion in powers of momentum~\cite{Bed.02}.
At present time, this framework is formulated only in a no-core shell model space,
that is currently limited to light systems which can not exhibit most of collective
exotic modes here discussed~\cite{Ste.09}.

Recently the Hartree-Fock plus RPA based on correlated realistic NN interaction
$V_{UCOM}$~\cite{Fel.98} has been established and employed in description 
of collective low-amplitude motion in nuclei~\cite{Paa.06,PPHR.06} and it has further
been extended toward SRPA~\cite{Pap.09}.
However, exotic structure phenomena in nuclei away from the valley of stability
still represent unsolved problem for theory frameworks based on low-momentum 
NN interactions $V_{low-k}$ and $V_{UCOM}$,
as well as inter-nucleon forces derived within the  $\chi$EFT~\cite{Epe.09}. 
As pointed out in modeling the excitation spectra of light nuclei, 
e.g. $^{10,11}$B, $^{12,13}$C, based on no-core shell model with NN and NNN 
interactions derived within $\chi$EFT, the inclusion of NNN force provides important
contribution and improves the theory results in comparison with experiment~\cite{Nav.08}.
The crucial open problem that remains to be solved, is to 
provide quantitative and simultaneous description of nuclear ground state properties
(e.g. binding energies, radii, nucleon separation energies, etc.) together with excitation
energies and transition strengths of giant resonances and exotic modes of excitation.
Within the existing theory frameworks, based on $V_{low-k}$ or $V_{UCOM}$, there
are difficulties to reach simultaneous quantitative description of excitation energies
of monopole, dipole and quadrupole giant resonances in stable nuclei,  and fine structure properties 
such as low-energy excitations in nuclei away from the valley of stability still remain
out of reach~\cite{Paa.06,Pap.09}. Since the nuclear many body theory based on 
realistic NN interactions also necessitates inclusion of the NNN force,
its implementation may improve the present status on description of collective modes
of excitation and could open perspectives for studies of exotic modes in nuclei away 
from the valley of stability.

\section{Multipole excitations in neutron-rich nuclei}

In recent years the PDR has been at the center of theoretical research on exotic modes
of excitation in nuclei with neutron excess~\cite{PVKC.07}, partly owing to profound 
progress in experimental studies, mainly 
based on photon scattering and Coulomb dissociation of radioactive ion beams, which 
resulted in valuable experimental data and opened new problems
in understanding the nature of low-energy excitations in nuclei with isospin
asymmetry. In the following, an overview is given for the critical questions
about the nature of PDR and exotic multipole excitations, including both theoretical 
and experimental aspects that are currently not fully understood. 

(i) {\it Are the low-energy dipole states attributed to the PDR collective?}

In contrast to giant resonances which are characterized by collective motion
of nucleons, there are contradictory results on collectivity and 
nature of the PDR~\cite{PVKC.07}. 
From the theory side, there is still no consensus whether the low-lying dipole strength 
could be attributed to a collective excitation mode. Whereas the relativistic
quasiparticle random phase approximation (RQRPA), by its amplitudes and
transition matrix elements, provides clear evidence about
degree of collective nature of the PDR~\cite{PVKC.07,Paa.09}, other self-consistent theory 
frameworks  provide rather limited evidence supporting the notion of
resonant structure of the PDR.

(ii) {\it Where is the exact location of the PDR excitation energy?}

Similar as in the case of giant resonances, the PDR centroid energy
decreases with the neutron excess, resulting in characteristic 
crossing between the PDR excitation energy and the neutron separation energy
for various isotope chains (e.g. Ni, Sn, and Pb)~\cite{PNVR.05}. 
For lower-mass isotopes the PDR is located 
below the neutron threshold, whereas in more neutron rich isotopes the PDR 
energy is higher than the neutron separation energy. Consequently, 
it is expected that there are important uncertainties in currently
available data on low-energy transition strength in nuclei with isospin asymmetry
where the PDR energy is close or above the neutron threshold.
Although there is a considerable amount of successful $(\gamma,\gamma')$ studies
(e.g. Refs.~\cite{Har.00, Zil.02, Sav.08,Sav.09,Sch.07,Wag.08,Rus.09}),
data on dipole excitations are available only below the neutron 
separation threshold, i.e. high-energy tail of the PDR may be missing.

(iii) {\it Can one resolve the missing E1 strength in photon scattering experiments?}

An important open problem in understanding the nature of PDR arises due to considerable 
amount of missing strength in $(\gamma,\gamma')$ measurements. 
The RQRPA calculations of the low-lying dipole strength along Sn isotope chain resulted
in significantly larger amount of B(E1) transition strength than experimental values from photon 
scattering~\cite{PVKC.07, Ozel.07}. Recent $(\gamma,\gamma')$ study confirmed
that quite a considerable amount of low-energy dipole transition strength could not 
be observed with existing experimental techniques and methods~\cite{Sav.08}, i.e.
the missing strengths can vary from a few percent to factor of three, depending on
the nucleus under consideration. Novel experimental facilities, e.g. the NEPTUN tagger facility~\cite{Sav.09},
could provide information not only about dipole transitions below the neutron threshold, 
but also across and above, and could possibly recover significant part of the 
missing low-energy dipole strength.

(iv) {\it Is there a connection between the PDR in stable and exotic nuclei?}

An important open problem related to the PDR is resolving the connection
between its nature in stable nuclei and those away from the valley of stability.
This connection may not be smooth in approaching the neutron drip-line,
where additional new structure phenomena may appear, e.g. di-neutron
vibration modes~\cite{Mat.05}. Although a general trend of increasing
pygmy dipole strength with neutron-proton asymmetry $\alpha=(N-Z)/A$
has been established~\cite{Kli.07}, there are uncertainties 
in establishing systematic dependence on $\alpha$.

Most of available experimental data on the low-lying dipole excitations 
related to the PDR in nuclei with isospin asymmetry are rather
close to the valley of stability. The pioneering experiment on the PDR in 
unstable nuclei, exploiting Coulomb dissociation of high-energy radioactive
beams, provided crucial information about low-energy dipole spectra
for $^{130,132}$Sn~\cite{Adr.05}.
However, due to the experimental techniques employed, the transition strength
is provided only at energies above the neutron separation energy which is
rather high, and some part of the PDR low-energy tail might be missing. Similar is
in the case of low-energy transition strength obtained in the $\gamma$ decay
from Coulomb excitation of $^{68}$Ni~\cite{Wie.09}. 

(v) {\it What is the macroscopic picture of nucleon vibrations in the PDR?}

Studies probing the nature of PDR result in contradicting results. 
Whereas the RPA transition densities indicate that the PDR corresponds to the skin
oscillation mode, where excess of weakly bound neutrons oscillate against the
proton-neutron core~\cite{PVKC.07}, the studies based on quasiparticle phonon
model showed that the PDR is characterized by vortical
motion of nucleons~\cite{Rye.02}. On the other hand,  relativistic RPA study of pronounced low-energy
states obtained for isoscalar dipole and toroidal operators resulted in vortical motion
in nucleon velocity fields as a genuine property
that is inherent for the low-lying dipole states~\cite{Vre.02}. Therefore, it is not fully
understood whether one could establish a connection between the vortical nature of low-lying dipole
states and those of the PDR, i.e. if there is a link between the neutron skin and
toroidal nuclear motion.

(vi) {\it What is the nature of the isoscalar-isovector splitting of the PDR?}

At present time, very little is known from the experimental side about the 
underlying structure of low-energy excitations, and available data mainly include 
excitation energies and B(E1) values for low-lying dipole states. Whereas transition densities
have been measured in the past for giant resonances, due to experimental difficulties in the case of 
PDR these are not available, and alternative methods are needed to understand how neutrons and
protons oscillate in the low-energy modes.
An important study along these lines, based on $(\gamma,\gamma')$ and $(\alpha,\alpha'\gamma)$
scattering, indicated that the structure of the low-energy excitations seems to
be rather complex, i.e. the lower part appears more sensitive to the isocalar probe,
whereas the higher part could be excited in both cases~\cite{Sav.06}.
The corresponding RQRPA study indicated that the more pronounced pygmy 
structure at lower energy is composed of predominantly isoscalar states with 
surface-peaked transition densities~\cite{Paa.09}. At somewhat higher energy
the calculated E1 strength is primarily of isovector character, as expected for the
low-energy tail of the giant dipole resonance. However, the double hump
structure of the low-lying strength could not be reproduced by self-consistent
model calculations, i.e. in the case of RQRPA this was possible only by
introducing the experimental single-particle levels. This opens another
important problem in the current implementations of the QRPA; mean field theory 
frameworks should be further developed in order to provide improved description of
single-particle spectra for the relevant states around the Fermi level.

(vii) {\it Are there other exotic multipole excitations in unstable nuclei?}

Although at present time the most prominent exotic mode in neutron-rich 
nuclei is the PDR, one could expect that other exotic modes could appear
in almost any multipolarity and isospin channel in nuclei away from the valley
of stability. In particular, there is
a whole set of open problems, related to discerning the conditions for
existence of exotic monopole, quadrupole and octupole modes of excitation
in nuclei with pronounced isospin asymmetry. Crucial question
to be resolved is whether loosely bound nucleons, with wave functions spatially 
extended far beyond radii characteristic for stable isotopes, could coherently
contribute to novel excitation modes of various multipolarities. 
In a recent study based on quasiparticle phonon model, quadrupole excitations
in neutron-rich Sn isotopes have been explored above the first low-lying collective
state and below the particle threshold~\cite{Tso.09}. It has been shown that
the structure properties of these states are closely related to the neutron excess, indicating
the existence of a skin oscillation quadrupole mode.  This result
has not been independently confirmed by self-consistent theory frameworks and
it represents an open problem to be addressed.

Self-consistent models currently available to conduct studies along
these lines, are usually formulated in the random phase approximation, which limits
the field of research to low-amplitude motion only. Therefore, additional efforts are
needed to develop new self-consistent theory frameworks based on modern effective
interactions, which could describe vibrations in nuclei without imposing limitations
on the vibration amplitude.

\section{Excitations in proton drip-line nuclei}

Nuclei close to the proton drip-line are characterized by unique structure
properties due to the weakly bound proton orbitals and the states which are
bound only because of the presence of the Coulomb barrier.
The proton drip line is much closer to the line of stability than the
neutron drip line. Bound nuclei with an excess of protons over neutrons can
be found only in the region of light Z $\leq$ 20 and medium
mass 20 $<$ Z  $\leq$ 50 elements. Due to the presence of the Coulomb barrier, nuclei close to the
proton drip line generally do not exhibit a pronounced proton skin, except 
for very light elements. Recent studies of proton rich medium-mass nuclei, 
based on RQRPA and continuum RPA, showed that low-energy dipole states
appear close to the proton drip line, which correspond
to the proton PDR~\cite{Paa.05,PPPW.05}.  The analysis of the 
proton and neutron transition densities showed that
proton PDR states correspond to the oscillation of the proton excess
against an approximately isospin-saturated core. The RQRPA amplitudes
of the pygmy states are composed of many proton two-quasiparticle
configurations, where dominant configurations include weakly bound
proton orbitals and also the states which are bound only because
of the presence of the Coulomb barrier.

Apart from a limited number of studies related to dipole transitions, excitation modes in proton
drip-line nuclei are at present time vastly unknown. These excitations are interesting not
only as unique structure phenomena at the limits of stability, but also due to possible 
astrophysical relevance, e.g., in rp-process. Two-proton decay has been for the first time
introduced by theory~\cite{Gol.60} and only recently it has been experimentally 
confirmed in the vicinity of the proton drip-line~\cite{Pfu.02}. On the other hand,
two-proton radiative capture, dominated by the E1 process, has been studied under
conditions of stellar environment~\cite{Gor.95,Gri.05}. However, the role of 
the proton PDR and its structure properties in two-proton capture rates still remains
an open problem, and it has only been studied using three-body model for two-proton halo
candidate $^{17}$Ne~\cite{Gri.06}. Self-consistent theoretical models based on modern effective 
interactions, which also include the role of the particle continuum, have not yet
systematically been included in modeling these reactions. 

\section{Charge-exchange modes and weak interaction rates}

Theoretical description of astrophysically relevant weak interaction of charged and neutral leptons
with nuclei crucially depends on modeling reaction transitions from the nuclear initial to 
excited states. 
One of the major open problems in nuclear physics applications in astrophysics, is construction
of fully self consistent microscopic theory framework based either on modern energy density functionals
or realistic NN interactions supplemented by the NNN force, that could provide quantitative
description of excitations contributing to weak interaction rates and allow extrapolations to 
unknown regions of nuclide map important for nucleosynthesis reactions, e.g. along the r-process path.
In the evolution of supernova core collapse, nuclear reactions involving weak interaction play
an essential role~\cite{Jan.07}. In particular, $\beta$-decay and electron capture in the mass
region of iron nuclei create neutrinos
and influence the number of electrons. Since neutrinos leave the star almost unhindered, in this way
the stellar core is cooling and keeping its entropy low~\cite{Bet.79}.  On the other hand, changing
the number of electrons determines the pressure in the star during collapse, and could have direct
impact on the dynamics of core collapse. Neutrino-nucleus reactions can also contribute to the
dynamics of the collapse and the subsequent explosion phase~\cite{Hax.88} and their modeling
necessitate self-consistent description of nuclear excitations induced by the neutrino interaction
with nuclei.

Microscopic models for description of astrophysically relevant excitations are  
based on nuclear shell model or QRPA. At present time, 
weak-interaction cross sections for target nuclei beyond the $pf$-shell, and those including transitions 
of higher multipolarities, cannot be systematically evaluated with large-scale diagonalisation
shell-model due to huge configuration spaces and the lack of a reliable effective interaction in other mass regions. 
Hybrid models have been introduced, which combine the shell-model Monte Carlo (SMMC) 
together with the RPA~\cite{LKD.01}, in order to also provide spin-dipole and other excitations 
of higher multipolarities needed, e.g., for description of neutrino-nucleus cross sections and electron capture
rates. In a recent application of the hybrid model in description of neutrino-nucleus reactions, the isobaric analogue state (IAS) and Gamow-Teller (GT) transition strengths are calculated in the shell-model with new Hamiltonian for $pf$ shell, while excitations of $J>1$ multipolarities are obtained from the RPA~\cite{Suz.09}. In modeling weak interaction rates, shell model can provide only IAS and GT excitations, whereas other excited states still
remain out of the reach due to limitations in the configuration space. Obviously, development of novel
theoretical frameworks is needed, that could in a systematic and consistent way provide all the 
necessary transitions throughout the nuclide map, including nuclei at the limits of stability. Although these objectives
can be achieved starting from the self-consistent mean field theory frameworks supplemented with the 
(quasiparticle) RPA~\cite{Fra.05,Paa.04,Lia.08,Lia.09}, in this way important correlations inherent in the shell 
model are not taken into account. 

At present time, the properties of charge-exchange excitations in exotic nuclei remain mainly
unknown both from theoretical and experimental perspective. Similar as in the case of 
excitations without charge exchange, one could, at least in principle, expect that
weakly bound nucleons could support formation of novel modes of excitations involving 
isospin flip. The evolution of the properties of basic excitation modes toward
drip lines still remains unknown  (e.g. spin-dipole excitations). Detailed knowledge about these
and other charge-exchange excitations of multipolarities up to J$\approx$6
is crucial in description of weak interaction rates, especially at somewhat higher energies of incoming 
lepton projectile (e.g. electron, neutrino, etc.)~\cite{Paa.08}. 
An open problem to be resolved is to establish a novel theory framework for self-consistent
description of the single particle properties, occupation probabilities and
charge-exchange excitations in a systematic way throughout the nuclide map, 
by including higher order correlations going beyond the $ph$ configurations inherent in RPA.

\section{Exotic modes of excitation in deformed nuclei}

Deformed nuclei represent a great challenge in the quest for novel modes of
excitation and their description necessitates framework where deformation
is explicitly taken into account. 
Only very recently, fully self-consistent theory frameworks for deformed nuclei
have been established, based on quasiparticle random approximation with
axial symmetry, formulated using Gogny effective force~\cite{Per.07}, as well as
Skyrme~\cite{Ken.08} and covariant~\cite{AR.08, AKR.09} 
energy density functionals. The QRPA  allows systematic description of
various multipole transitions throughout the map of nuclides, including
nuclei near the drip lines. As pointed out in Ref.~\cite{Ken.08}, these approaches have
practical drawback due to necessity to include a large two-quasiparticle cutoff
which constrains the configuration space, and therefore computing times and 
memory storage are high. However, implementation of parallel codes~\cite{AR.08}
and on-going development of computing facilities will allow the configuration space 
to be considerably enlarged for the purposes in the quest for novel modes of excitation in
deformed nuclei.

The evolution of the low-lying
strength in deformed nuclei is determined by the interplay of isospin asymmetry and deformation;
greater neutron excess increases the total low-lying strength while deformation results in spreading
and hindering of the transition strength. 
In a model of a deformed nucleus in which protons and neutrons
have been described as interacting rigid rotors with axial symmetry, magnetic-dipole
collective state describing scissor-like rotational oscillations of protons against neutrons have
been predicted thirty years ago~\cite{Lud.78}. The scissors mode has been a topic in numerous theoretical studies,
and it has been experimentally observed in many nuclei by scattering of protons, electrons
or photons~\cite{Boh.84,Kne.96}. However, excitations in deformed nuclei away from the valley
of stability are mainly unknown and present an open problem for future studies. In fact,
only very recently the model calculations indicated that in neutron-rich $^{154}$Sm a 
magnetic dipole state of unique structure properties appears at low excitation energy 
at 2.5 MeV~\cite{Art.09}. It has been shown that this peak corresponds
to a novel exotic type of resonance with pronounced isoscalar character in which the 
neutron skin oscillates in a scissor like motion against the proton-neutron core. Apart from a few 
recent studies, excitations in exotic nuclei with deformation are at present time mainly unknown. 
These modes, both in neutron- and proton-rich nuclei, present a challenge not only for
both relativistic and nonrelativistic energy density functionals, but also for models
based on realistic NN and NNN interactions.

\section{Excitations in nuclei at finite temperature in stellar environment}

Nuclei at finite temperature are characterized by thermal population
of single-particle states and modifications of nuclear structure properties,
resulting in new features in excitation spectra. In general, description of
open-shell nuclei necessitates a consistent treatment of pairing correlations
together with finite temperature effects like, for instance, in the finite
temperature HFB+QRPA framework~\cite{Som.83}.
The phase transition from a superfluid to a normal state occurs at temperatures 
$T \approx 0.5-1$ MeV~\cite{Kha.04}, whereas for temperatures above 
$T \approx 4$ MeV contributions from states in the continuum become
large, and additional subtraction schemes have to be implemented to remove
the contributions of the external nucleon gas~\cite{Bon.84}.

Multipole responses in hot nuclei
have been explored in a variety of theoretical frameworks, e.g. RPA based on
schematic interactions~\cite{Civ.84}, linear response theory~\cite{Fab.83},
self-consistent Hartree-Fock plus RPA based on Skyrme functionals~\cite{Sag.84,PCKV.09},
and covariant energy density functionals~\cite{Yif.09}. The decay of hot nuclei
and thermal damping of giant resonances have also been studied in great 
detail~\cite{Gal.85,Bor.86}.
Whereas the main giant resonance peaks remain weakly affected at finite
temperatures up to T=3 MeV, long low-energy tails have been obtained.

In view of all the previous sections in this article, crucial topic that is
currently not understood is the question how various exotic modes of
excitation evolve with increased temperature, in particular under conditions
of the stellar environment where these modes could play significant role
in astrophysical processes (cf. Sec. 6). Since the major effect
due to finite temperature is expected in the low-energy part of the excitation
spectra, one could anticipate considerable modifications  in the properties 
of exotic modes of excitation or novel transition strength may appear~\cite{Yif.09}.

Particular challenge is self-consistent description of structure properties
of excitation modes in nuclei at the limits of stability at temperatures
in stellar conditions. In addition to weak binding of outermost nucleons,
finite temperature further redistributes the occupation numbers of 
single-particle states, resulting in additional weakening of the nuclear 
binding and modifications in excitation spectra.
At temperatures below the phase transition from a superfluid to normal state
($T < 0.5-1$ MeV), the structure properties will depend
on the interplay between the effects of the pairing correlations and finite
temperature, resulting in partial occupations of relevant single-particle
states around the Fermi level, which contribute to the excitation configuration 
space.

While the finite temperature effects on giant resonances are rather moderate, 
it is not understood to what extent finite temperature would support the
existence of exotic modes, or what are conditions for these modes to
become suppressed. Whereas some of these issues could be addressed
by self-consistent theory frameworks based on modern energy density
functionals, the knowledge from the experimental side is mainly
limited to giant resonances~\cite{Bra.89,Bau.98,San.06}, and considerable
efforts are necessary
to provide some information about the structure properties of weakly
bound nuclei at finite temperature. At present time, even very basic
information needed for description of weak interaction rates is not available
from experiment, namely the evolution of $\beta^+$ Gamow-Teller transition 
strength with temperatures in the range characteristic for presupernova 
conditions of stars (e.g. T=0-2 MeV). 
Although $GT^{\pm}$ transitions dominate $\beta$-decays and electron 
captures under supernova core-collapse conditions~\cite{Jan.07}, forbidden transitions
could also result in non-negligible contributions in astrophysical applications,
as shown for instance in the case of $\beta$-decays of nuclei near the r-process
path~\cite{Bor.03}. Therefore, forbidden transitions contributing to the weak interaction
rates should be revised within self-conistent theory frameworks at finite 
temperature, as well as their role in astrophysical applications. 

\section{Conclusion}

At present time, both theoretical and experimental studies of exotic modes
of excitation are to a large extent focused on discerning the nature of pygmy dipole
resonances. As discussed in Sec.~3, due to considerable progress in theoretical and 
experimental research within recent years, several questions 
are opened on the underlying structure properties of the PDR, including the
issues of collectivity, location of the PDR with respect to neutron threshold
energy, the missing strength from photon scattering experiments, connection between the
PDR in stable nuclei and those at the frontiers of stability, macroscopic picture
of nuclear vibrations, and isoscalar-isovector splitting of the PDR.
On the other side, nuclei at extreme isospin and finite temperature, as well
as deformed nuclei with isospin asymmetry, open perspectives for a
whole new class of monopole, dipole, quadrupole and octupole excitation modes,
as well as exotic charge-exchange vibrations. Since nuclei at finite temperature and
away from the valley of stability participate in astrophysically relevant reactions,
new knowledge on exotic nuclear modes of excitation may result in important
consequences in understanding the processes in stellar environment.
In the quest for novel exotic modes of excitation, development of theoretical 
frameworks plays a central role. Reliable microscopic description of excitations
necessitates fully self-consistent theory, able to include the effects
of the particle continuum and couplings to complex configurations, in order to
provide description of both escape and spreading widths.
These could be formulated by using modern energy density functionals constrained by the
strong interaction physics of QCD and/or by the phenomenological input. On the other
side, exotic modes of excitation also represent a great challenge and valuable test for theory
frameworks based on realistic NN interactions, including low-momentum
NN interactions,  $V_{low-k}$ and $V_{UCOM}$, as well as chiral NN
interactions, supplemented by three body forces.

\leftline{\bf ACKNOWLEDGMENTS} 
This work was supported by the Unity through 
Knowledge Fund (UKF Grant No. 17/08) and Ministry of Science, Education and Sports of the
Republic of Croatia (project No. 1191005-1010) and Croatian National Foundation for Science.



\end{document}